\documentclass[preprint,12pt]{elsarticle}
\usepackage{setspace}
\usepackage[]{natbib}
\usepackage{graphicx,amssymb,amsmath,times}
\usepackage{subfigure}
\usepackage{lscape}
\journal{New Astronomy}
\begin{document}
\begin{frontmatter}
\title{Search for TeV $\gamma$--ray emission from blazar 1ES1218+304 with TACTIC telescope during March-April 2013}
\author[label1]{K K Singh\corref{corr}}
\ead{kksastro@barc.gov.in}
\author[label1]{K K Yadav} \author[label1]{A K Tickoo} \author[label1]{R C Rannot} \author[label1]{P Chandra} 
\author[label1]{N K Agarwal} \author[label1]{K K Gaur} \author[label1]{A Goyal} \author[label1]{H C Goyal} 
\author[label1]{N Kumar} \author[label1]{P Marandi} \author[label1]{M Kothari} \author[label1]{H Bhatt} 
\author[label1]{K Chanchalani} \author[label1]{N Chouhan} \author[label1]{V K Dhar} \author[label1]{B Ghosal} 
\author[label1]{S R Kaul} \author[label1]{M K Koul} \author[label1]{R Koul} \author[label1]{K Venugopal} 
\author[label1]{C K Bhat} \author[label1]{C Borwankar} \author[label2]{J Bhagwan} \author[label2] {A C Gupta}
\address[label1]{Astrophysical Sciences Division, Bhabha Atomic Research Cente. \\
Mumbai- 400 085, India.}
\address [label2]{Aryabhatta Research Institute of Observational Sciences. \\
Nainital- 263129, India.}
\begin{abstract}
In this paper, we present results of  TeV $\gamma$--ray observations of the high synchrotron peaked BL Lac object 
1ES 1218+304 (z=0.182) with the  $TACTIC$ (TeV Atmospheric Cherenkov Telescope with Imaging Camera). The observations 
are primarily motivated  by the unusually hard  GeV-TeV spectrum of the source despite its relatively large  redshift. 
The source is observed in the TeV energy range with the $TACTIC$ from March 1, 2013 to April 15, 2013 (MJD 56352--56397) 
for a total observation time of 39.62 h and no evidence of TeV $\gamma$--ray activity is found from the source. The 
corresponding  99$\%$ confidence level upper limit on the integral flux above a threshold energy of 1.1 TeV is estimated 
to be 3.41 $\times10^{-12}$ photons cm$^{-2}$ s$^{-1}$ (i.e $<23\%$ Crab Nebula flux) assuming a power law differential 
energy spectrum with photon index 3.0, as previously observed by the $MAGIC$  and $VERITAS$ telescopes. For the study of 
multi-wavelength emission from the source, we use nearly simultaneous optical, UV and and X--ray data collected by the 
UVOT and XRT instruments on board the \emph{Swift} satellite and high energy $\gamma$--ray data collected by the Large 
Area Telescope on board the \emph{Fermi} satellite. We also use radio data at 15 GHz from OVRO 40 m telescope in the 
same period. No significant increase of activity is detected from radio to TeV $\gamma$--rays from 1ES1218+304 during 
the period from March 1, 2013 to April 15, 2013.
\end{abstract}
\begin{keyword}
Blazars: 1ES 1218+304, Cherenkov Imaging  telescope: Very high energy gamma-rays, Multi-wavelength observations.
\end{keyword}

\end{frontmatter}

\section{Introduction}  
Blazars are observed to emit highly variable non-thermal radiation spanning the entire electromagnetic spectrum. The Spectral 
Energy Distribution (SED) of blazars is assumed to be dominated by emission from a relativistic jet pointing  close to 
our line of sight. The jets in blazars consist of ultra-relativistic particles embedded in a magnetic field, with the entire 
plasma flowing outward from the central region with relativistic speed. The characteristic SED of blazars shows two broad 
non-thermal well defined continuum peaks. The first hump peaks somewhere between the infrared (IR) and X--ray bands, whereas 
the second hump exhibits a maximum at $\gamma$--ray energies. The origin of low energy peak is attributed to  the synchrotron 
emission of relativistic electrons in the magnetic field of the jet. In leptonic  models \cite{LM92,CD93,MS94,MB00}, the 
$\gamma$--ray emission in High Energy (HE: E$>$100 MeV) and Very High Energy (VHE: E$>$100 GeV) regimes is attributed to 
the inverse Compton (IC) scattering of low energy photons by the same electron population producing the synchrotron radiation 
(SSC: Synchrotron Self Compton) or with the  possible contribution from external photons (EC: External Compton). In hadronic 
models \cite{MB92,Ah00,MP01,MU03}, both electrons and protons are accelerated to ultra-relativistic energies, with protons 
exceeding the threshold for photo-pion production on the soft photon field in the jet. In these models, the HE and VHE 
$\gamma$--ray emissions are dominated by proton synchrotron emission, neutral pion decay photons, synchrotron and Compton 
emission from secondary decay products of charged pions.
\par
Blazars are broadly classified in two groups namely BL Lacertae objects (BL Lacs) and Flat Spectrum Radio Quasars (FSRQs) 
\cite{Urry95}. BL Lacs are characterized by the weak or absence of thermal features like broad emission lines in their optical 
spectra. On the other hand, FSRQs exhibit luminous broad emission lines in their optical spectra. The peak frequency of the 
synchrotron component in the SED of blazars is usually used to subdivide them in 3-classes: low-synchrotron peaked 
(LSP: $\nu_{peak}<10^{14}$ Hz), intermediate-synchrotron peaked (ISP: $10^{14}$ Hz $<$ $\nu_{peak}$ $<$ $10^{15}$ Hz) and 
high-synchrotron peaked (HSP: $\nu_{peak}>10^{15}$ Hz) blazars \cite{Abdo10}. BL Lacs are assumed to be good VHE candidates 
for ground based TeV telescopes because their IC peak is in the TeV regime. To date, BL Lacs observed at VHE energies are 
predominantly high frequency peaked objects, which have been predicted as possible TeV candidate blazars \cite{Cost02}.   
\par
Blazars often show violent flux variability from radio to VHE $\gamma$--rays at different time scales which may or may not be 
correlated. Therefore, simultaneous multi-wavelength (MWL) observations are important to understand the underlying physics 
of blazars. Furthermore, VHE observations of blazars at cosmological distances would help in constraining the intensity 
and spectrum of Extragalactic Background Light (EBL), which is a very important physical quantity to understand the structure 
and star formation history \cite{Steck92}. Thus, besides their importance for study of emission mechanisms 
and relativistic jet dynamics, growing interest for blazar study is also  motivated by the use of VHE spectra as a probe for 
EBL \cite{Ah06}.
\par
1ES 1218+304  (RA:$12^h 21^m 23.5^s$, Dec :$30^{\circ} 10^{\prime} 43.2^{\prime\prime}$) is an HSP BL Lac object located at a 
redshift z = 0.182 \cite{Bade98}. The source was discovered as a candidate BL Lac object on the basis of its X-ray emission and 
was identified as X-ray source 2A 1219+30.5 \cite{Wils79,Ledd81}. Being an X--ray bright source, it was predicted to be a TeV 
candidate blazar from the position of the synchrotron peak in its SED \cite{Cost02}. 
\par
The $HEGRA$ (High Energy Gamma Ray Astronomy) collaboration observed 1ES 1218+304 during 1996-2002 for 3.9 h and reported an upper 
limit of 12$\%$ of Crab nebula flux  above 750 GeV \cite{hegra03}. The VHE emission from 1ES 1218+304 was first detected by the 
$MAGIC$ (Major Atmospheric Gamma ray Imaging Cherenkov) telescope in 2005 above  120 GeV \cite{Alber06}. The differential energy 
spectrum (d$\Phi$/dE=f$_0$ E$^{-\Gamma}$), with power law index ($\Gamma$) of 3.0$\pm$0.4, was also reported by the $MAGIC$ group. 
In May 2006, 1ES 1218+304 was the target of $HESS$ (High Energy Stereoscopic System) observation campaign and these observations did 
not yield  any statistically significant signal from the source \cite{hess08}. The corresponding 99.9$\%$ Confidence Level (CL) limit 
on integral flux above 1 TeV  was  reported  to  be  17$\%$ of the $HESS$ Crab Nebula flux. The $STACEE$ (Solar Tower Atmospheric 
Cherenkov Effect Experiment) detector monitored 1ES 1218+304 during 2006-2007 above 160 GeV but did not detect any significant 
$\gamma$--ray emission from the source \cite{Stace11}. In 2007, $VERITAS$ (Very Energetic Radiation Imaging Telescope Array System) 
telescope observed VHE emission above 160 GeV from the source at a persistent level of 6$\%$ of the Crab Nebula flux \cite{Accia09}. 
The differential energy spectrum of the source was found to be compatible with a power law of index 3.08$\pm$0.34. Using the 
lower limit EBL from galaxy counts \cite{MP00, Faz04}, the intrinsic power law index was found to be 2.32$\pm$0.37. During the
December 2008 - May 2009 monitoring campaign of 1ES 1218+304, $VERITAS$ telescope revealed a prominent flaring activity from the source 
at 20$\%$ of the Crab Nebula flux above 200 GeV \cite{Accia10}. The time averaged differential energy spectrum, in the energy range 
0.2-1.8 TeV, was found  to  be a power law with index of 3.07$\pm$0.09. This flaring activity was characterized  by a variability time 
scale of days. The corrected spectrum after accounting for absorption due to EBL suggests a very hard intrinsic source spectrum with 
index $\le$ 1.28$\pm$0.28 \cite{Kren08} in VHE regime based on lower limit EBL model \cite{LW08}.   
\par
\emph{Fermi}-LAT (Large Area Telescope) also detected significant emission from 1ES 1218+304 during its first two years of operation \cite{Nol12}. 
The \emph{Fermi}-LAT spectrum of the source is described by a power law with index 1.71$\pm$0.07, making it one of the hardest spectrum 
sources in MeV-GeV range. Due to relatively large redshift of the source, we expect a significant attenuation of TeV photons above 
1 TeV due to EBL absorption. This makes the source a good candidate for probing EBL using TeV observations. It is because of these 
characteristics of the source we were motivated to observe the source with $TACTIC$ (TeV Atmospheric Cherenkov Telescope with Imaging Camera) 
in the TeV energy region.  
\par
The paper is organized as follows. Section 2 gives a brief description of the $TACTIC$ telescope and in Section 3 we present 
details of observation and data analysis. In Section 4 we present the  MWL data analysis of the source from radio to HE. In 
Section 5 we discuss the results and the conclusions are presented in Section 6. 

\section{TACTIC Telescope}
The $TACTIC$ $\gamma$--ray telescope is located at Mount Abu (1300 m asl, $24.6^{\circ}$N, $72.7^{\circ}$E), India \cite{Tact07}. 
The telescope is equipped with a F/1-type tessellated light collector of approximately 9.5 $m^{2}$ area consisting of 34 front-face 
aluminium coated, spherical glass mirror facets of 60 cm diameter. The  point-spread  function  has a full width at half maxima 
(FWHM) of 0.185$^{\circ}$ ($\equiv$12.5mm)  and D$_{90}$  $\sim$0.34$^{\circ}$ ($\equiv$22.8mm). Here, D$_{90}$ is   
defined as the diameter of a circle, concentric with the centroid of the image, within which 90$\%$  of reflected 
rays lie. The telescope deploys a  349-pixel imaging camera, with a uniform pixel resolution of $0.31^{\circ}$ and 
a $5.9^{\circ}\times5.9^{\circ}$ field-of-view, to record  atmospheric Cherenkov events produced by an incoming 
cosmic-ray particle or a $\gamma$-ray photon. Data  used  in this work  have  been  collected  with   inner 
225 pixels  where the innermost 121 pixels (11 $\times$ 11 matrix) are used for generating the event trigger. 
The trigger scheme is based on a Nearest Neighbour Non-collinear Triplets trigger criterion. Apart from  
generating  the  prompt trigger with  a coincidence  gate width  of $\approx$18ns, the trigger generator has a 
provision for producing a chance coincidence output based on  $^{12}$$C_{2}$  combinations from various groups 
of closely spaced 12 channels. The data acquisition and control system of the telescope  is designed around a 
network of PCs running the QNX (version 4.25) real-time operating system. The triggered events are digitized by 
CAMAC based 12-bit charge to digital converters (CDC), which have a full scale range of 600 pC. The relative gain 
of the photomultiplier tubes is monitored regularly  once in 15 minutes by flashing a blue LED, placed at a 
distance of about 1.5m from the camera. Other details regarding  hardware and software features of the  data 
acquisition and control system of the telescope are discussed in \cite{TDAS}.
\par
Major upgrade, involving replacement of signal and high voltage cables and installation of new Compound Parabolic 
Concentrators (CPC) was taken up in  November-December 2011 for  improving  the  sensitivity  of the  telescope. While  
detailed simulation and  experimental results on the Crab Nebula, after upgrade  will be presented  elsewhere, we present 
here only a  brief   summary of the  upgrade  work. New  CPCs with square entry and  circular exit aperture were installed 
on the $TACTIC$ imaging  camera  in  order to increase its  photon collection  efficiency. The collection efficiency of 
the new CPCs was measured to be about 80$\%$ in  the wavelength range 400--550 nm. Apart from removing the dead space 
in between the photomultipliers substantially, the new CPCs has also helped us to improve the  gamma/hadron 
segregation capability of the telescope. In addition, the  trigger criteria was also modified by including more nearest 
neighbor collinear triplet combinations. A dedicated CCD  camera  was  also installed for conducting detailed point run 
calibrations  and data collected were successfully used for determining the position of the source in the  image  plane 
with an accuracy better than  $\pm$ 3 arc min. The point run calibration data (i.e.  position  of the star image in  
the camera) were also incorporated in the  analysis software so that appropriate corrections can be applied in an off 
line manner while calculating source  position  dependent  image   parameters. The analysis  procedure  was  improved 
by using the Asymmetry parameter so that additional hadronic  background can be further removed by identifying the ``head/tail'' 
feature of Cherenkov images. It is worth mentioning here that the $\gamma$--ray images have their head 
closer to the assumed source position in the imaging camera and  thus  can  be selected preferentially by imposing  
Asymmetry $>$0 cut.
\par
The upgrade of the telescope has led to an increase in the prompt coincidence rate from 2.33 Hz to 4.70 Hz close to the 
zenith. Furthermore, as a result of the upgrade the $\gamma$--ray rate from  the Crab  Nebula  has also increased from  
9.13 h$^{-1}$ to 15.30 h$^{-1}$ and this translates to the reduction in the telescope threshold energy from 1.2 TeV to  
0.87 TeV.  
 
\section{TACTIC Observations and Results}
The blazar 1ES 1218+304 was monitored with the $TACTIC$ from March 1, 2013 to April 15, 2013 (MJD 56352--56397)
at zenith angles between $6^{\circ}$ and $45^{\circ}$. All the data were collected in tracking mode, during which the 
telescope continuously monitored the source. This mode of observation maximizes the source observation time 
and increases the possibility of detecting flaring activity from the source. About 54 h of data were collected  
during 18 nights of observations as per the details summarized in Table 1. 
\begin{table}[h]
\caption{Summary of $TACTIC$ observations of 1ES 1218+304}
\begin{center}
\begin{tabular}{lcccc}
\hline
Month		&Observation dates			&Observation Time (h)	&Selected data (h)\\
\hline
March 2013	&3,4,8,9,10,12,13,14,17,18,19		& 26.00			&16.79 \\	
April 2013	&6,7,9,10,11,12,13			& 28.00			&22.83 \\
\hline
Total 		&18 Nights				&54.00			&39.62\\
\hline
\end{tabular}
\end{center}
\end{table}
\subsection{TACTIC Data Analysis}
Apart  from  excluding the observations during bad atmospheric conditions, several standard data quality tests have been applied 
to the raw  data for selecting  clean  data for further analysis. The data quality checks include the following: 
compatibility   of  the   prompt coincidence rates with the expected zenith angle behavior, Poission distribution for arrival 
times of prompt events and  steady behavior of chance coincidence rates with time. After applying the above data quality checks 
the final data sample reduces to  $\sim$ 39.62 h. For gamma/hadron separation we  have followed the standard Hillas parameter 
analysis \cite{Hill85} where  each  Cherenkov  image  is  characterized by its moments. In this procedure 
each Cherenkov image is characterized by various image parameters like \emph{length} (L), \emph{width} (W), \emph{distance} (D), 
\emph{alpha} ($\alpha$), \emph{size} (S), \emph{frac2} (F2) and \emph{asymmetry} (ASYM). This technique  was later refined 
to Dynamic Supercuts  procedure where  $S$ dependent  shape parameters  of the image as well as its orientation were used 
for segregating the $\gamma$--rays from the background cosmic--rays \cite{Hill98,Moh98}. The $\gamma$--ray selection 
criteria used in the  present  work are given in Table 2. These cuts have been optimized using 25 h of actual observation 
data on the Crab Nebula during November 2012. When applied to the remaining data on the Crab Nebula, the above cuts yield 
consistent detection of a $\gamma$--ray signal at a sensitivity level of  N$_{\sigma}$ $\sim$ 1.40 $\sqrt{T}$ 
(where T is the  observation time in hours).
\begin{table}[h]
\caption{ Dynamic Supercuts selection  criterion used for analyzing the $TACTIC$ data.}
\begin{center}
\begin{tabular}{lclc}
\hline
Parameters  &  Cuts Value \\ 
\hline
L   &  $0.11^{\circ}$ $\le$ L $\le$ (0.1000 + 0.0520 $\times$ ln S)$^{\circ}$ \\ 
W   &  $0.06^{\circ}$ $\le$ W $\le$ (0.0850 + 0.0160 $\times$ ln S)$^{\circ}$ \\
D   &  $0.50^{\circ}$ $\le$ D $\le$ (1.27 $\times$ cos$^{0.95}$$\theta$)$^{\circ}$ ($\theta$=zenith angle) \\ 
S   &  $\ge$  310 dc (digital counts)\\ 
$\alpha$  &  $\alpha$ $\le$ 18$^{\circ}$ \\ 
F2  &  $\ge$  0.35 \\ 
L/W &  $\ge$ 1.55 \\
ASYM & $\ge$ 0.0\\
\hline
\end{tabular}
\end{center}
\end{table} 
\par
A well established  procedure to extract the $\gamma$--ray signal from the cosmic ray background is to plot the 
frequency distribution of  $\alpha$ parameter (defined as the angle between the major axis of the image and the 
line between the image centroid and camera center) of shape (L,W) and D selected events. This distribution is expected to 
be flat  for the isotropic background of cosmic events. For $\gamma$--rays, coming from a point source, the distribution 
is expected to show a peak at  smaller $\alpha$ values. Defining $\alpha$ $\leq$ 18$^{\circ}$ as the $\gamma$--ray domain 
and 27$^{\circ}$ $\leq$ $\alpha$ $\leq$ 81$^{\circ}$ as the background region, the number of $\gamma$-ray events is then 
calculated  by subtracting the expected number of background events (calculated on the basis of background  region) from 
the  $\gamma$--ray domain events. The statistical significance of $\gamma$--ray like events is calculated using the methodology 
proposed by Li and Ma \cite{LiMa83}. 
\par
In order to validate the proper functioning of the telescope and the data analysis methodology, we  are collecting data 
on the Crab Nebula regularly. Although, during the observing season 2012-2013, the Crab  Nebula was  observed  with  
$TACTIC$ right from   November 2012  onwards, we present here, as a representative example, the results  of our observations 
for approximately 11.33 h only, which were carried out from March 1, 2013  to March 13, 2013. The main purpose of doing this 
is to compare the results of the Crab Nebula observations with that of 1ES1218+304. There is an overlap of 6 nights when both 
the Crab Nubula and 1ES1218+304 have been observed one after the other. Figure 1(a) gives the $\alpha$--distribution when the 
data collected  on the Crab Nebula for 11.33 h is analyzed. The events selected after using the Dynamic Supercuts procedure 
yield an excess of  169$\pm$32 $\gamma$-ray like events with  a statistical significance of 5.47$\sigma$. The corresponding 
$\gamma$--ray rate turns  out to be (14.91$\pm$2.82)h$^{-1}$. Since the average  zenith of 28$^{\circ}$ for 11.33 h data on 
the Crab Nebula is close to the average  zenith  of  21$^{\circ}$ for 39.62 h data on 1ES1218+304, one can express the 
$\gamma$--ray rate observed from 1ES1218+304 in Crab Unit (CU:1 CU$\equiv$14.91$\pm$2.82 h$^{-1}$). Figure 1(b) gives the 
$\alpha$--distribution for the data collected on 1ES 1218+304 for 39.62 h. The analysis of the data yields an excess of  
2 $\pm$ 56 $\gamma$--ray  events  with  a statistical significance of 0.04$\sigma$, which suggests that there is no evidence for 
a $\gamma$--ray signal from the source during the period  of our  observations. The source is thus possibly 
in a low state which is below the sensitivity level of the $TACTIC$. The upper limit estimation on the integral VHE 
$\gamma$--ray flux from the source is  described below.
\subsection{Upper Limit Calculation for TACTIC}
Using  the  probability density  function  of the number of excess events we have determined the  upper limit on the excess events 
(N$_{UL}$) by using the  methodology  proposed  by  Helene \cite{Hel83} and  the method involves solving the following 
equation for N$_{UL}$, 
 \begin{equation}\label{eq1}
 \beta I\left(\frac{ -N_{exc}}{\sigma_{exc} }\right)=  I\left(\frac{ N_{UL} -N_{exc}}{\sigma_{exc} }\right)
  \end{equation}
where (1-$\beta$)$\times$ 100$\%$ is the confidence level, N$_{exc}$ is  number of  excess events  with   $\sigma_{exc}$ 
as its standard deviation. The  function I(x) is  given by 
\begin{equation}\label{eq2}
I(x)=  \frac{1}{\sqrt{2\pi}} \int^{\infty}_{x} \; e^{-t^2/2} \; dt=  \frac{1}{2} \;erfc( \frac{x}{\sqrt{2}}) 
\end{equation}
Where  erfc(x) is the complementary error function. On solving equation 1 with N$_{exc}$= 2, $\sigma_{exc}$= 56 (refer $\alpha$--plot 
shown in Figure 1(b)) and $\beta$=0.01, we get N$_{UL}$ $\approx$ 146 as 99$\%$ confidence level upper limit on the excess events from the 
source. Knowing that the $\gamma$--ray rate of (14.91$\pm$2.82)h$^{-1}$ corresponds to 1 CU, the resulting 99$\%$ limit on the rate 
of excess events from  1ES1218+304  translates to 3.68h$^{-1}$ (i.e 146/39.62h or 0.25 CU). Alternatively, on dividing the 
upper limit on the excess events  by the  product of effective collection area $\sim 3.0\times 10^8$ cm$^{2}$ (obtained after applying the 
Dynamic Supercuts) and observation  time $\sim$ 39.62 h, we find the 99$\%$ upper limit on the integral flux to be 
3.41$\times10^{-12}$ photons $cm^{-2} s^{-1}$ above threshold energy of 1.1 TeV. The values of effective collection area and threshold energy 
used above correspond to zenith angle  of $25^{\circ}$. If we assume  a source spectrum  similar to that of the Crab Nebula 
(i.e d$\Phi$/dE = 2.79$\times$ 10$^{-11}$ E$^{-2.59}$ cm$^{-2}$ s$^{-1}$ TeV$^{-1}$ ; as measured  by the $HEGRA$ group \cite{Crab}) and 
also found to match very well with the spectrum obtained from the $TACTIC$ \cite{Tact07}, the above 99$\%$ limit translates to an integral 
flux upper limit of 0.26 CU. Referring back to the upper limit, if we assume a steeper spectrum (i.e d$\Phi$/dE $\propto$ E$^{-3.0}$) for 
1ES1218+304, similar  to the one observed  by $MAGIC$ and $VERITAS$ telescopes \cite{Alber06,Accia09}, the above upper limit on the integral 
flux corresponds to about 0.23 CU above a threshold energy of 1.1 TeV. The differential flux upper limit at 99$\%$ confidence level is found to 
be 6.2$\times$ 10$^{-12}$ photons cm$^{-2}$ s$^{-1}$ TeV$^{-1}$.
 
\section{Analysis of multi-wavelength  data}
The MWL data for 1ES 1218+304  were collected in optical and UV by \emph{Swift}-UVOT, in X-rays by \emph{Swift}-XRT, and in HE $\gamma$--rays 
by \emph{Fermi}-LAT during March 1, 2013 to April 15, 2013 (MJD 56352--56397). The details of observation and data reduction for these instruments 
are described below. 
\subsection{\emph{Fermi}-LAT data}
The \emph{Fermi}-LAT is a pair conversion $\gamma$--ray telescope  sensitive to photon energy in the MeV-GeV regime \cite{Lat09}. The instrument 
has been designed to measure the directions, energies and arrival times of incident photons in the energy range 20 MeV to $>$ 300 GeV while 
rejecting the background from cosmic rays. In its nominal scanning mode, it surveys the whole sky every 3 h with a large field of view of about 
2.4 steradian. The \emph{Fermi}-LAT data for 1ES 1218+304 were retrieved from the publicly available NASA data 
base\footnote{{http://fermi.gsfc.nasa.gov/ssc/data/access}} during the period March 1, 2013 to April 15, 2013 (MJD 56352--56397).  We selected the 
good quality events from the ``source class'' over the energy range 100 MeV--100 GeV and the events were extracted from a circular region of 
interest (ROI) with radius $15^{\circ}$ centered at the source position (RA=$12^h 21^m 23.5^s$,  Dec=$30^{\circ} 10^{\prime} 43.2^{\prime\prime}$). 
In addition, we excluded the events observed with zenith angles $>$ 100$^{\circ}$ to limit contamination from Earth limb $\gamma$--rays, and events 
detected while the spacecraft rocking angle was $>$ 52$^{\circ}$ to avoid time intervals during which the bright limb of the Earth entered the LAT 
field of view. 
\par 
The data obtained in this manner were analyzed using the standard \emph{Fermi} Science-Tools software package (version v9r27p1). 
We used P7$\_$SOURCE$\_$V6 instrument response function with the galactic and isotropic diffuse emission models 
gal$\_$2yearp7v6$\_$v0.fits and iso$\_$p7v6source.txt. All the point sources from \emph{Fermi}-LAT second source catalog 
(2FGL) \cite{Nol12} within $20^{\circ}$ of 1ES 1218+304, including the source of interest itself were considered in source 
model file. Sources within the ROI were fitted with power law models with the normalization and spectral index as free 
parameters, while those beyond ROI had their model parameters frozen to those as reported in second source catalog \cite{Nol12}. 
An unbinned likelihood spectral analysis was performed to produce the light curve with the standard analysis tool 
\emph{gtlike} implemented in Science-Tools software package. Since the source is not always detected  at high statistical 
significance, we have produced the five day binned light curve with minimum statistical significance accepted for each time 
bin as TS $\ge$ 4, where TS is the test statistic defined as twice the difference of the log(likelihood) with and without 
the source respectively \cite{Mat96}. The time averaged spectrum of the source was obtained by fitting a power law model with 
the normalization as free parameter and spectral index set to the value obtained by integrating the data over entire period of 
observation in the energy range 100 MeV-100 GeV. The details of the source spectrum obtained in the present work from 
\emph{Fermi}-LAT observations are described in Section 5.2.  
\subsection{\emph{Swift}-XRT \& UVOT data}
During March 1, 2013 to April 15, 2013 (MJD 56352--56397) only six days of observations are available from \emph{Swift} with 
XRT (X--Ray Telescope), covering the 0.3-10 keV energy band \cite{xrt05}, and UVOT (UV/Optical Telescope), covering 180-600 nm 
wavelength range \cite{uvot05}. 
\par
\emph{Swift}-XRT data were reduced following the standard procedure\footnote{http://www.swift.ac.uk/analysis/xrt/} 
using $FTOOLS$. The data were collected in window timing (WT) mode for all the observations. The task {\sc xselect} (ver V2.4b) 
within the {\sc HEASoft} package (v6.13) with recent calibration files (ver. 20120209) was used to analyse the data. The spectra 
and light curves of the source were extracted using a circular region with radius of $23^{\prime\prime}$ around the source. The 
spectra and light curves of nearby background region were extracted within an annulus with inner radius of $24^{\prime\prime}$ and 
outer radius of $45^{\prime\prime}$ around the source. The corresponding exposure maps and ancillary response files (ARFs) were generated 
using the tasks {\sc xrtexpomap } and {\sc xrtmkarf} for all the observations, respectively. The spectra were binned using 
{\sc grppha} to ensure a minimum of 20 counts per bin to perform the $\chi^2$ minimization for fitting the spectrum with model 
$phabs*zpowerlaw$ and the fluxes were calculated using {\sc cflux}. The best-fit parameters given in Table 3 were derived for 
individual observation with a neutral hydrogen column density fixed to its Galactic value 1.99$\times10^{20}$ $cm^{-2}$ obtained 
from NASA/IPAC Extragalactic Database (NED)\footnote{ned.ipac.caltech.edu}. The best fit average parameters were also derived 
using simultaneous fitting of the spectra of six observations.
\par
The source 1ES 1218+304 was observed with \emph{Swift}-UVOT using all filters (V, B, U, UVW1, UVM2, UVW2) in image mode over 
six days during $TACTIC$ observations. The image mode level II data of all the filters were used in the present analysis with 
latest calibration files of UVOT \cite{Breev11}. The data were processed with the standard 
procedure\footnote{http://www.swift.ac.uk/analysis/uvot/} using {\sc uvotmaghist} task of {\sc heasoft} package. The UVOT 
source counts were extracted from a $5^{\prime\prime}$ sized circular region centered on the source position, while the 
background was extracted from a nearby larger, source free, circular region of $10^{\prime\prime}$ radius. 
The observed magnitudes were converted into fluxes using conversion factors given in \cite{Pool08}. The observed magnitudes obtained 
in six filters during the period from March 1, 2013 to April 15, 2013 are reported in Table 4. 
\begin{table}
\begin{minipage}{\textwidth}
\caption{Spectral analysis of \emph{Swift}-XRT data during March 1, 2013 to April 15, 2013  using power law spectrum with neutral hydrogen
 density fixed at 1.99$\times10^{20}$ $cm^{-2}$.}
\label{Table:XRT}
\begin{center}
\begin{tabular}{ccccc}
\hline
MJD    & Obs-ID           & {Photon index}   	       &{Flux (10$^{-11}$ erg cm$^{-2}$ s$^{-1}$)}          &$\chi^2_{red}(dof)$\\
\hline
56360   &sw00030376021    & $2.05_{-0.04}^{+0.04}$     	& $4.89_{-0.06}^{+0.06}$		     &1.23 (48)\\        
56366   &sw00030376022    & $1.94_{-0.04}^{+0.04}$     	& $6.49_{-0.09}^{+0.08}$		     &0.90 (51)\\    
56369   &sw00030376023    & $2.11_{-0.03}^{+0.04}$     	& $5.84_{-0.06}^{+0.06}$		     &0.80 (64)\\   
56387   &sw00030376024    & $2.23_{-0.05}^{+0.05}$     	& $4.03_{-0.06}^{+0.06}$		     &0.93 (43)\\
56390   &sw00030376025    & $2.13_{-0.04}^{+0.05}$     	& $4.42_{-0.06}^{+0.06}$		     &1.43 (46)\\
56396   &sw00030376026    & $2.23_{-0.04}^{+0.04}$     	& $4.31_{-0.05}^{+0.05}$		     &1.38 (58)\\
\hline
\end{tabular}
\end{center}
\end{minipage}
\end{table}
\begin{table}
\begin{minipage}{\textwidth}
\caption{Summary of \emph{Swift}-UVOT observed magnitudes during March 1, 2013 to April 15, 2013 in six filters.}
\label{Table:UVOT}
\begin{center}
\begin{tabular}{ccccccc}
\hline
MJD       &{V} 		   &{B}		   &{U}	   	   &{W1}	   &{M2}	     &{W2}\\
\hline
56360     &16.15$\pm$0.08  &16.78$\pm$0.06 &15.86$\pm$0.05 &15.85$\pm$0.05  &15.59$\pm$0.04  &15.83$\pm$0.03\\        
56366     &16.43$\pm$0.01  &16.98$\pm$0.07 &15.96$\pm$0.05 &16.06$\pm$0.05  &16.30$\pm$0.07  &16.24$\pm$0.04\\    
56369     &16.27$\pm$0.08  &16.87$\pm$0.06 &16.01$\pm$0.05 &15.91$\pm$0.05  &15.79$\pm$0.05  &15.80$\pm$0.03\\   
56387     &16.30$\pm$0.08  &16.76$\pm$0.06 &15.95$\pm$0.05 &15.85$\pm$0.05  &15.79$\pm$0.05  &15.87$\pm$0.04 \\
56390     &16.17$\pm$0.08  &16.86$\pm$0.06 &16.10$\pm$0.05 &15.89$\pm$0.05  &15.76$\pm$0.05  &15.84$\pm$0.04 \\
56396     &16.21$\pm$0.07  &16.79$\pm$0.05 &15.82$\pm$0.04 &15.76$\pm$0.04  &15.72$\pm$0.04  &15.88$\pm$0.03\\
\hline
\end{tabular}
\end{center}
\end{minipage}
\end{table}
\subsection{OVRO data}
The $OVRO$ (Owens Valley Radio Observatory) is a fast cadence 15 GHz radio telescope with diameter of 40 m \cite{ovro11}. 
The telescope is a f/0.4 type parabolic reflector on an alt-azm mounting system and is equipped with dual beamed off axis optics 
and a cooled receiver installed at prime focus. The source 1ES 1218+304 was observed at 15 GHz using $OVRO$ 
telescope\footnote{www.astro.caltech.edu/ovroblazars/data} for six days during $TACTIC$ observations as part of \emph{Fermi} MWL 
blazar monitoring program and data of these observations are used in the present work.    
\section{Results and Discussion}
\subsection{Light curve analysis}
The MWL light curve of 1ES 1218+304 observed by various instruments during March 1, 2013 to April 15, 2013 (MJD 56352--56397)
is shown in the Figure 2. Since no significant $\gamma$--ray emission has been detected from the source with the $TACTIC$,  
we have shown the 99$\%$ CL upper limit on integral flux above 1.1 TeV for $TACTIC$ observations in Figure 2(a).
The five-day binned light curve of the source observed with \emph{Fermi}-LAT during the same period is presented in Figure 2(b).
All the points reported in the Figure 2(b) correspond to TS $>$ 4. From the figure, it is evident that there is no statistically 
significant variation in the $\gamma$--ray activity in the energy range 100 MeV-100 GeV during $TACTIC$ observations and the 
average flux level during this period is found to be (4.17$\pm$0.82)$\times10^{-8}$ photons cm$^{-2}$ s$^{-1}$. It is important 
to mention here that the source has been categorized as highly variable on monthly time scale in  the  second \emph{Fermi} 
catalog \cite{Nol12} and thus some enhanced activity may be quite consistent with its past behavior. 
\par
Nearly simultaneous X--ray light curve observed by \emph{Swift}-XRT for six days of monitoring of the source is depicted 
in Figure 2(c). From the figure, we observe that the soft X--ray emission from the source is consistent with the average flux 
level (5.01$\pm$0.34)$\times10^{-11}$ erg cm$^{-2}$ s$^{-1}$, except for some enhanced activity on March 15, 2013 (MJD 56366). 
During the enhanced activity on March 15 in X-rays, the HE activity is also observed to be slightly higher with respect to the 
average flux. We have also looked into the archival X--ray data in the energy bands 2-20 keV and 15-50 keV from 
\textit{MAXI}\footnote{{http://maxi.riken.jp/top/index.php}} and 
\textit{Swift}-BAT\footnote{{http://heasarc.nasa.gov/docs/swift/results/transients}} instruments respectively and these data do 
not show any detection above 3$\sigma$. This indicates that during the period of $TACTIC$ observations, the source was not active 
in hard X-ray regime. The  simultaneous UV (W1,M2,W2 filters) and optical(V,B,U filters) light curves of the source for six days 
of observations are presented in Figure 2(d) and 2(e) respectively. The UVOT flux points included in the light curve have not 
been de-reddened. No unusual activity is observed in the source with \emph{Swift}-UVOT instrument during the period of $TACTIC$ 
observations. The radio observations at 15 GHz available for six days are shown in Figure 2(f). No significant variations are 
observed in radio emission from the source during this period. 
\subsection{Spectral Analysis}
The spectral energy distribution of 1ES1218+304  during March 1, 2013 to April 15, 2013 using broad band data discussed above 
is shown in Figure 3. For VHE $\gamma$--rays we plot the data from the observations with $MAGIC$ and $VERITAS$ telescopes along 
with the 99$\%$ confidence level upper limit on integral flux obtained from $TACTIC$ in the present work. The flux points 
reported by $MAGIC$ group are based on the observations carried out during the period January 9-15, 2005 \cite{Alber06}. 
During six days of observations, no flux variability on timescales of days was found and the time averaged spectrum was 
described by a power law with $\Gamma=3.0\pm0.4$. The $VERITAS$ flux  points taken from \cite{Accia09} correspond  to observations 
during January-March 2007 for a total observation time of 17.4 h. The time averaged differential spectrum was described by a power law  
with photon index 3.08$\pm$0.34 and the integral flux  above  200 GeV  was 6$\%$ of the Crab Nebula flux. It is evident from the figure 
that the measured flux points are consistent with each other in the overlaping energy regime of the two telescopes. The VHE flux points 
measured with $MAGIC$ and $VERITAS$ telescopes as well as upper limit from $TACTIC$ have been corrected for EBL absorption using the mean 
level density model proposed by Franceschini et al. (2008) \cite{Franc08}. 
\par
The LAT data points have been  obtained by dividing the energy range 0.1--100 GeV into four energy bands: 0.1--1 GeV, 
1--10 GeV, 10--20 GeV and 20--100 GeV. The time averaged GeV spectrum measured by \emph{Fermi}-LAT during this period is described 
by a power law with normalization factor f$_0$=(8.47$\pm$1.20)$\times$ 10$^{-9}$ cm$^{-2}$ s$^{-1}$ GeV$^{-1}$ and photon index 
$\Gamma=1.78\pm0.06$. The HE photon index $\Gamma$=1.78$\pm$0.06 obtained in the present study is consistent with the value reported 
from quiescent state monitoring of 1ES 1218+304 by \emph{Fermi}-LAT \cite{Abdo09a,Abdo09b}. The photon index observed by \emph{Fermi}-LAT 
during the first two years of observation of 1ES 1218+304 is 1.709$\pm$0.067 \cite{Nol12}. 
\par
The \emph{Swift}-XRT flux points have been obtained from simultaneous fitting of the spectra of six observations in three energy 
bands:0.3-1 keV, 1-2.5 keV and 2.5--6 keV using {\sc cflux}. Beyond 6 keV, \emph{Swift}-XRT observations are not statistically 
significant to perform spectral analysis. The X-ray flux points measured with XRT as shown in Figure 3 have been corrected for 
Galactic absorption using a neutral hydrogen column density of 1.99$\times10^{20}$ $cm^{-2}$ obtained from NED\footnote{ned.ipac.caltech.edu}. 
The time averaged soft X--ray spectrum measured with XRT during $TACTIC$ observations is described by a power law with photon index 
2.13$\pm$0.01. The X--ray emission level of (1.93$\pm$0.01)$\times$ 10$^{-11}$ erg cm$^{-2}$ s$^{-1}$ in the energy range 2-10 keV, 
observed in our present study is below the flux level (2.64$\pm$0.02)$\times$ 10$^{-11}$ erg cm$^{-2}$ s$^{-1}$ obtained from  
\emph{XMM-Newton} measurements during 2001 observations \cite{Blust04}. 
\par
The simultaneous optical and UV emission in the wavelength range 180-600 nm as measured by \emph{Swift}/UVOT instrument in all six 
filters (V,B,U,W1,M2,W2) is shown in the Figure 3. The flux densities in all filters are estimated from the dereddened magnitudes 
with galactic absorption $A_v=0.056$ and $R_v=A_v/E(B-V)=3.1$ \cite{Schla11}, using the methodology proposed in \cite{Pool08}. The 
UVOT flux points have been obtained by multiplying the mean of flux densities with the bandpass (FWHM) of the corresponding filter 
\cite{Pool08}. Error in the mean density is obtained through the standard error propagation method. The radio flux point at 15 GHz 
obtained from OVRO telescope data archive\footnote{http://www.astro.caltech.edu/ovroblazars/data/data.php} corresponds to the mean 
emission level from the source during $TACTIC$ observations.  
\par
The broadband data points presented in Figure 3 indicate that the SED of the source can be described by two humps: first one peaking at 
X--ray energies and second at GeV energies. This implies that the MWL emission from 1ES 1218+304 can be possibly compared with the  
predictions of SSC model for blazar emission, but detailed SED modeling of the source is beyond the scope of this work. 
\par
Tang et al. (2010) have reproduced the SED of 1ES 1218+304 with the inhomogeneous jet model and the homogeneous SSC model \cite{Yun10}. 
They emphasize that the leptonic model is very successful in explaining multi-band emissions from the  source and also point out that 
the VHE $\gamma$--ray data from the $MAGIC$ and $VERITAS$ telescopes can be fitted with the strict 
lower-limit EBL model. Using the \emph{Swift}, $MAGIC$ and $VERITAS$ telescopes data the SED of the source has been modeled in 
\cite{SED10} by employing a time-dependent SSC code for obtaining  the physical parameters of the emission region. The short-time 
variability of the source has also been studied by Weidinger \& Spanier (2010) \cite{Weidinger} for reproducing the  light curve 
observed by $VERITAS$ telescope. They suggest that the  light curve can be reproduced by assuming a changing level of electron 
injection compared to the constant state.  

\section{Conclusions}
Our $\gamma$--ray observations of 1ES 1218+304 (z=0.182) with the  $TACTIC$ from March 1, 2013 to April 15, 2013 (MJD 56352-56397) 
for a total observation time of $\sim$39.62 h do not show any evidence of TeV $\gamma$--ray signal from the source. The  MWL data 
in the X-ray and HE bands, as measured  by \emph{Swift}-XRT (0.3-10 keV) and \emph{Fermi}/LAT (0.1--100 GeV) 
respectively, do not reveal any unusual activity from  the source. The optical and UV emissions observed with \emph{Swift}-UVOT
instrument and radio observations at 15 GHz with 40 m $OVRO$ telescope during $TACTIC$ observations also do not indicate any 
flaring activity from the source. It is important to  point out here that, because of the variable nature of blazars in general, the VHE  
emission  from  1ES1218+304  may  increase significantly  during future flaring episodes and may even easily exceed the limit 
reported in the present work. Hence  the  upper limit  presented  here  only constrains  the flux  during our observation period. 
Our blazar observation program will continue to monitor this  source.   
    
\section*{Acknowledgment}
We thank the anonymous reviewer for his/her suggestions which improved the quality of the paper. 
The authors would like to convey their gratitude to all the concerned colleagues of the  Astrophysical Sciences  
Division  for their contributions towards the instrumentation, observation and analysis  aspects of the $TACTIC$ telescope.  
We acknowledge the useful discussions held with N.G. Bhatt on various aspects of data analysis. We acknowledge the use 
of public data obtained through \textit{Fermi} Science Support Center (FSSC) provided by NASA. This work made use of data 
supplied by the UK \emph{Swift} Science Data Centre at the University of Leicester. This research has made use of the 
NASA/IPAC Extragalactic Database (NED) which is operated by the Jet Propulsion Laboratory, California Institute of Technology, 
under contract with the National Aeronautics and Space Administration. 

\newpage
\begin{figure}
\begin{center}
\includegraphics[width=0.85\textwidth,angle=0]{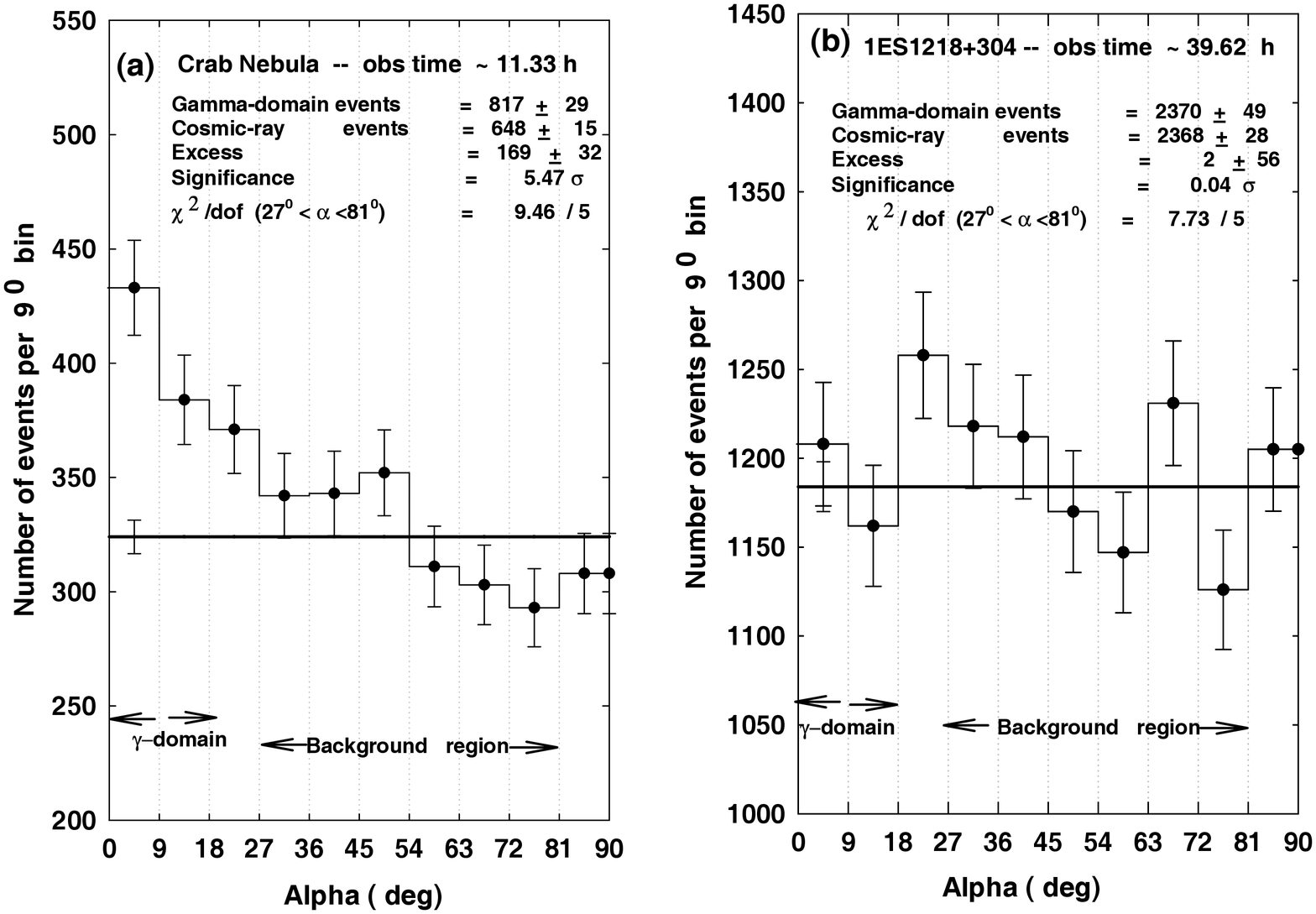}
\caption{(a) On-source alpha plot of Crab Nebula for 11.33 h of data collected during March 1, 2013  to March 13, 2013 
(b) On-source alpha plot of 1ES1218+304 for 39.62 h of data collected during March 1, 2013  to April 15, 2013 (MJD 56352--56397). 
The horizontal lines  in these figures indicate  the expected  background  in the $\gamma$--domain  obtained by using the background 
region with 27$^\circ$ $\leq$ $\alpha$ $\leq$ 81$^\circ$.}
\end{center}
\end{figure}

\begin{figure}
\begin{center}
\includegraphics[width=1.0\textwidth]{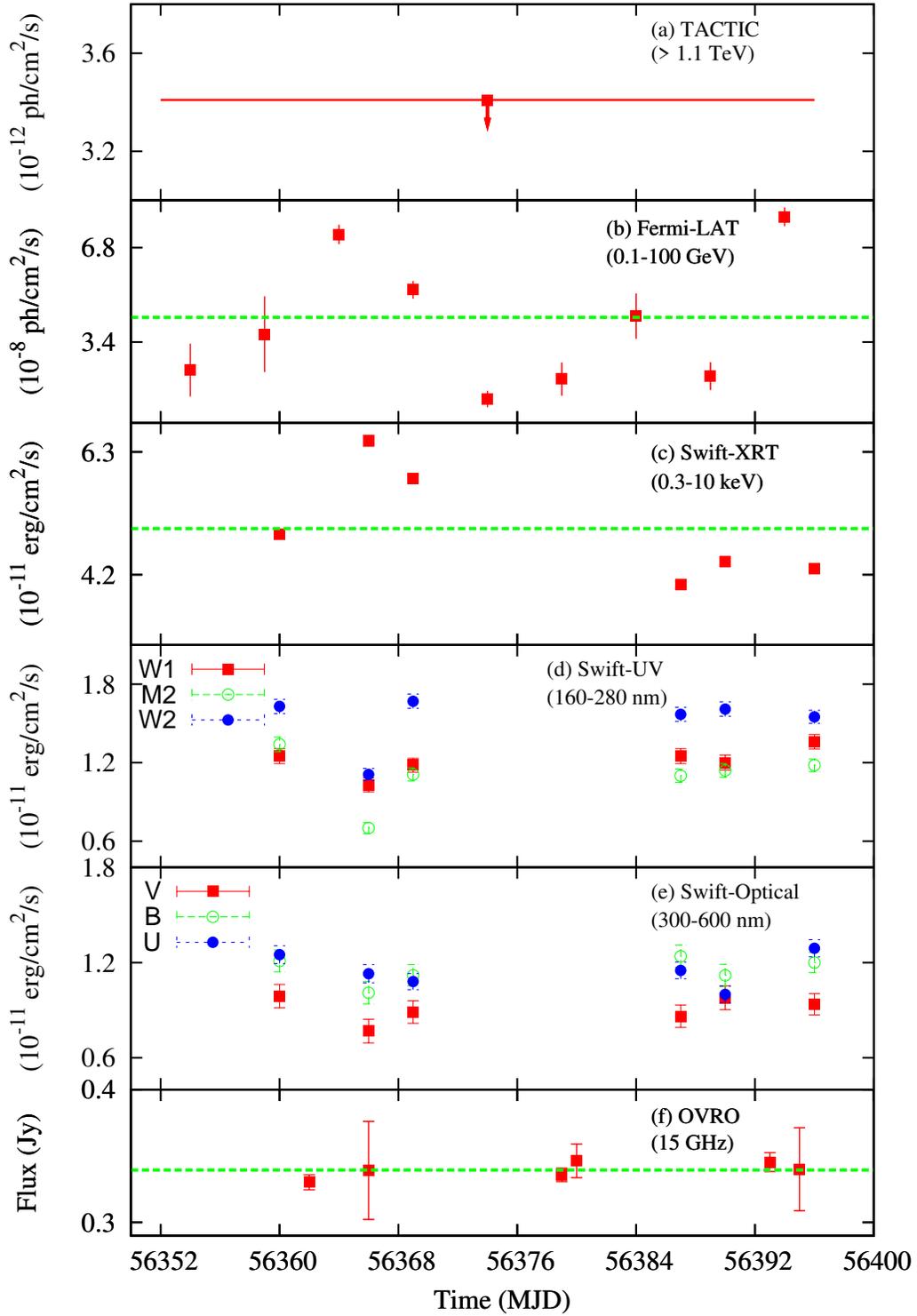}
\caption{ Multi-wavelength light curve of 1ES 1218+304 during March 1, 2013 to April 15, 2013 covering VHE $\gamma$--rays by 
$TACTIC$, HE $\gamma$--rays by \emph{Fermi}-LAT, X--rays by \emph{Swift}-XRT, UV/optical by \emph{Swift}-UVOT instruments and 
radio by $OVRO$ telescope. 99$\%$ confidence upper limit in VHE $\gamma$--rays by $TACTIC$ is indicated as downward arrow. Each 
point in the 5 day binned light curve of \emph{Fermi}-LAT in HE $\gamma$--rays corresponds to TS $>4$. X--ray, UV/Optical and 
radio light curves represent daily flux values for 6 days of observations available during the monitoring of the source with $TACTIC$. 
The horizontal dotted lines (shown in (b), (c) \& (f)) represent the average emission level in the respective energy bands.}
\end{center}
\end{figure}

\begin{figure}
\begin{center}
\includegraphics[width=0.8\textwidth,angle=-90]{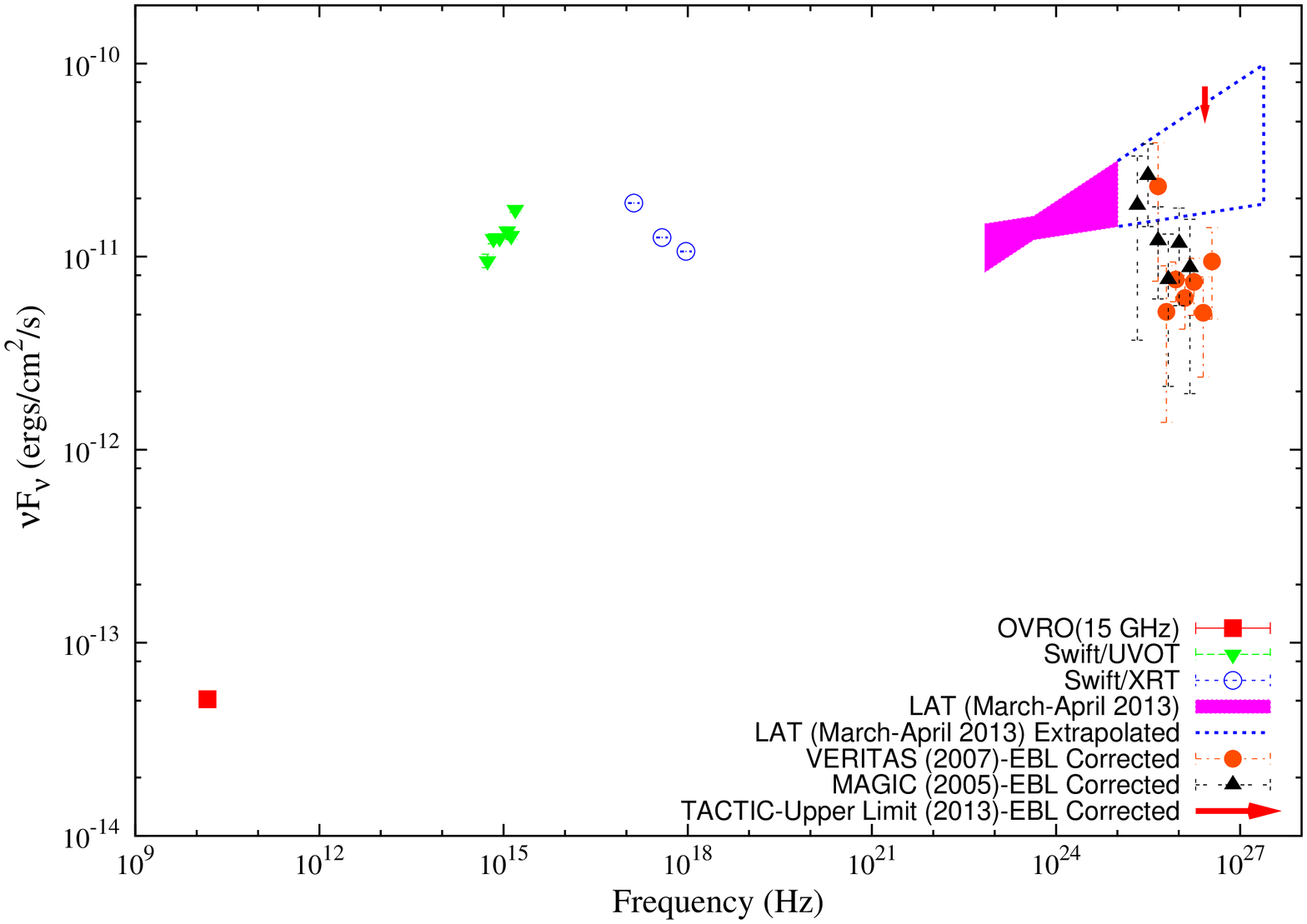}
\caption{Multi-wavelength SED of 1ES1218+304 measured during March 1, 2013 to April 15, 2013. VHE points correspond to the fluxes 
measured by: $MAGIC$ \cite{Alber06}, $VERITAS$ \cite{Accia09} telescopes along with $TACTIC$ 99$\%$ flux upper limit at $\sim$1.1 TeV 
(present work). All the VHE flux points have been corrected for EBL absorption using Franceschini et al. (2008) model \cite{Franc08}. 
\emph{Fermi}-LAT spectrum measured during March-April 2013 is shown as butterfly (filled curve) and same has been extrapolated in the 
$TACTIC$ energy range 0.87--10 TeV (dotted lines). \emph{Swift}-XRT points are time averaged fluxes for six days of observations. 
\emph{Swift}-UVOT points correspond to time averaged fluxes measured in all filters (V,B,U,W1,M2,W2). The time averaged radio flux at 
15 GHz is taken from the OVRO observations.}
\end{center}
\end{figure} 

\begin{thebibliography}{}
\bibitem{LM92}    L. Maraschi et al. 1992, ApJ, 397, L5	
\bibitem{CD93}    C. Dermer and R. Schleickeiser, 1993, ApJ, 416, 458 
\bibitem{MS94}    M. Sikora et al., 1994, ApJ, 421, 153		  
\bibitem{MB00}    M. Blazejowski et al., 2000, ApJ, 545, 107	  
\bibitem{MB92}    K. Mannheim and P. L. Biermann, 1992, A\&A, 253, L21 
\bibitem{Ah00}    F. Aharonian et al., 2000, New Astron., 5, 377     
\bibitem{MP01}    A. Mucke and R. J. Protheroe, 2001, Astropart. Phys., 15, 121 
\bibitem{MU03}    A. Mucke et al., 2003, Astropart. Phys., 18, 593	
\bibitem{Urry95}  C. Urry and P. Padovani, 1995, PASP, 107, 803 
\bibitem{Abdo10}  A. A. Abdo et al., 2010, ApJ, 716, 30        
\bibitem{Cost02}  L. Costamante and G. Ghisellini, 2002, A\&A, 384, 56 
\bibitem{Steck92} F. W. Stecker et al., 1992, ApJ, 390, L49  
\bibitem{Ah06}	  F. Aharonian et al., 2006, Nature, 440, 1018	
\bibitem{Bade98}  N. Bade et al., 1998, A\&A, 334, 459        
\bibitem{Wils79}  A. S. Wilson et al., 1979, MNRAS, 187, 109 
\bibitem{Ledd81}  J. E. Ledden et al., 1981, ApJ, 243, 47    
\bibitem{hegra03} M. Tluczykont et al., 2003, $28^{th}$ ICRC, 2547 
\bibitem{Alber06} J. Albert et al., 2006, ApJ, 642, L119     
\bibitem{hess08}  F. Aharonian et al., 2008, A\&A, 478, 387  
\bibitem{Stace11} C. Mueller et al., 2011, Astropart. Phys., 34, 674 
\bibitem{Accia09} A. Acciari et al., 2009, ApJ, 695, 1370    
\bibitem{MP00}    P. Madau and L. Pozzetti, 2000, MNRAS, 312, L9 
\bibitem{Faz04}   G. G. Fazio et al., 2004, ApJS, 154, 39     
\bibitem{Accia10} A. Acciari et al., 2010, ApJL, 709, L163   
\bibitem{Kren08}  F. Krennrich et al., 2008, ApJ, 689, L23    
\bibitem{LW08}	  L. Levenson and E. Wright, 2008, ApJ, 683, 585 
\bibitem{Nol12}   P. L. Nolan et al., 2012, ApJS, 199, 31    
\bibitem{Tact07}  R. Koul et al., 2007, NIM A, 578, 548      
\bibitem{TDAS}    K. K.Yadav  et al., 2004, NIM A, 527, 411  
\bibitem{Hill85}  A. M. Hillas et al., 1985, Proc. 19th ICRC, 3, 445 
\bibitem{Hill98}  A. M. Hillas et al., 1998, ApJ, 503, 744   
\bibitem{Moh98}	  G. Mohanty et al., 1998, ApJ, 9, 15	      
\bibitem{LiMa83}  T. P. Li and Y. Q. Ma, 1983, ApJ, 272, 317   
\bibitem{Hel83}   O. Helene, 1983, NIM A, 212, 319    
\bibitem{Crab}	  F. A. Aharonian et al., 2000, ApJ, 539, 317   
\bibitem{Lat09}   W. B. Atwood et al., 2009, ApJ, 697, 1071  
\bibitem{Mat96}	  J. R. Mattox et al., 1996, ApJ, 461, 396   
\bibitem{xrt05}	  D. N. Burrows et al., 2005, Space Sci. Rev. 120, 165 
\bibitem{uvot05}  P. W. A. Roming et al., 2005, Space Sci. Rev. 120, 95
\bibitem{Breev11} A. A. Breeveld et al., 2011, AIPC, 1358, 373 	   
\bibitem{Pool08}  T. S. Poole et al., 2008, MNRAS, 383, 627       
\bibitem{ovro11}  J. L. Richards et al., 2011, ApJS, 194, 29     
\bibitem{Franc08} A. Franceschini et al., 2008, A\&A, 487, 837   
\bibitem{Abdo09a} A. A. Abdo et al., 2009, ApJ, 700, 597      
\bibitem{Abdo09b} A. A. Abdo et al., 2009, ApJ, 707, 1310      
\bibitem{Blust04} A. J. Blustin, M. J. Page and G. B. Raymont, 2004, A\&A, 417, 61 
\bibitem{Schla11} E. F. Schlafly and D. P. Finkbeiner, 2011, ApJ, 737, 103  
\bibitem {Yun10}   Y. Tang, Z. Dai and L. Zhang, 2010, RA\&A 10, 415
\bibitem{SED10}  M. Ruger et al., 2010, MNRAS, 401,973          
\bibitem{Weidinger} M. Weidinger and F. Spanier, 2010, A\&A, 515, A18
\end{thebibliography}
\end{document}